\documentclass[11pt,a4paper]{article}

\usepackage{fullpage}
\usepackage{amsmath}
\usepackage{amssymb}
\usepackage{graphicx}
\usepackage[all,poly,curve,knot,cmtip]{xy}
\usepackage{citesort}

\newcommand {\be} {\begin{eqnarray*}}
\newcommand {\ee} {\end{eqnarray*}}
\newcommand {\bea} {\begin{eqnarray}}
\newcommand {\eea} {\end{eqnarray}}

\newcommand{\tsent}[2]{d{#1}/d{#2}}

\title{An exact solution of the five-dimensional Einstein equations with four-dimensional
de Sitter-like expansion}

\author{\textbf{Tom$\acute{\mbox{a}}\check{\mbox{s}}$ Liko}${}^{\dagger}$ and
\textbf{Paul S. Wesson}${}^{\ddagger}$ \\
\\ {\small \it Department of Physics, University of Waterloo} \\
{\small \it Waterloo, Ontario, Canada, N2L 3G1}}

\begin{document}

\maketitle

Correspondence:

${}^{\dagger}$tliko@astro.uwaterloo.ca

${}^{\ddagger}$Mail address above; phone: (519) 885-1211 ext. 2215; fax: (519) 746-8115;

Webpage: http://astro.uwaterloo.ca/${}^{\sim}$wesson

\begin{abstract}
We present an exact solution to the Einstein field equations which is
Ricci and Riemann flat in five dimensions, but in four dimensions is a
good model for the early vacuum-dominated universe.
\end{abstract}

\hspace{0.3cm}\textbf{PACS}: 04.20.Jb, 04.50.+h, 98.80.Cq

\section{Introduction}

There has recently been an uprising in interest in finding exact solutions of the Kaluza-Klein
field equations in five dimensions (5D) which reproduce and extend known solutions of the Einstein
field equations in four dimensions (4D) \cite{ledbel,agubel,bellini,clx,liu}.  Particular interest
revolves around solutions which are not only Ricci flat ($R_{AB}=0;A,B,\ldots\in\{0,1,2,3,4\}$ where
$R_{AB}$ is the 5D Ricci tensor), but also Riemann flat ($R_{ABCD}=0$, where the vanishing of the
Riemann-Christoffel tensor means that we are considering the analog of the Minkowski metric in 5D)
\cite{acm1,liuwes1,wesliu,mcmanus,dsv,leo1}.  This is because it is possible to have a flat 5D manifold
which contains a curved 4D submanifold, as implied by Campbell's embedding theorem
\cite{campbell,magaard,rrt,rtz,lrtr,seawes,anderson}.  So, the universe may be ``empty''
and simple in 5D, but contain matter of complicated forms in 4D \cite{wesleo,wesson1}.  (This idea
has been extended to higher-dimensional manifolds that are not Ricci-flat, in particular manifolds
with non-zero cosmological constant \cite{andlid,dahrom1}, scalar field sources \cite{adlr}, as well
as manifolds with an arbitrary non-degenerate Ricci tensor \cite{dahrom2}.  In addition, the
Campbell-Magaard theorem has been used to study the embedding of Randall-Sundrum-type branes in 5D
manifolds \cite{dahrom3}, suggesting that the curvature of any given brane is not necessarily
determined by its stress-energy content.)

Despite the physical appeal of this idea, it is mathematically non-trivial to realize.  Solutions of
the flat and empty Einstein equations in 5D which correspond to solutions of
$G_{\alpha\beta}=T_{\alpha\beta}$ ($\alpha,\beta,\ldots\in\{0,1,2,3\}$) in 4D with acceptable physics,
are rare.  (Here $G_{\alpha\beta}$ is the 4D Einstein tensor and $T_{\alpha\beta}$ is the induced
stress-energy tensor obtained via the standard reduction of the 5D equations to their 4D
counterparts; see reference \cite{wesson1}.  We use units throughout which render the speed of light
and Newton's gravitational constant invisible via $c=1,8\pi G=1$.)  In what follows, we present and
derive the properties of an exact 5D solution which provides a good 4D model for the vacuum-dominated
early universe.

\section{A New Solution and its Properties}

Consider the five-dimensional line element with coordinates $t,r,\theta,\phi,\ell$ such that
\bea
dS^{2} &=& \frac{\ell^{2}}{L^{2}}dt^{2}
         - \left[\ell\sinh\left(\frac{t}{L}\right)\right]^{2}d\sigma_{3}^{2}
         - d\ell^{2}\nonumber\\
d\sigma_{3}^{2} &=& \left(1+\frac{kr^{2}}{4}\right)^{-2}
                  (dr^{2}+r^{2}d\theta^{2}+r^{2}\sin^{2}\theta d\phi^{2})
\quad
\mbox{and}
\quad
k = -1.
\label{manifold}
\eea
In five dimensions this defines a manifold $(\mathcal{M},g_{AB})$ that is indeed both Ricci-flat
and Riemann-flat, thus giving Minkowski space $\mathbb{M}^{5}$ in a different coordinate system.
That (\ref{manifold}) satisfies the Ricci-flat equations $R_{AB}=0$ may be shown by tedious algebra
(e.g. using the equations of reference \cite{wesson1}), and confirmed by computer (e.g. using the
program GRTensor of reference \cite{lakmus}).  The only humanly-practical way to show that
(\ref{manifold}) also satisfies the Riemann-flat equations $R_{ABCD}=0$ is by computer, as may be
verified.

The physical properties of the matter associated with (\ref{manifold}) may, again, be derived either
analytically or computationally.  The basic procedure, in either approach, is to separate the purely
4D terms in $R_{AB}=0$ from the other ones, compare with $G_{\alpha\beta}=T_{\alpha\beta}$, and thereby
obtain $T_{\alpha\beta}=T_{\alpha\beta}(x^{4},\partial g_{AB}/\partial x^{C})$.  Since the Einstein
equations $G_{AB}=0$ in empty 5D are equivalent to $R_{AB}=0$ by straight algebra, what we are doing
here is simply solving in effect the 5D Einstein equations, comparing the results to the 4D Einstein
equations, and thereby evaluating the stress-energy tensor $T_{\alpha\beta}$ necessary to balance
the latter set of equations.

This procedure has in recent years been much used.  A review of the algebraic technique and a list
of applications is available \cite{wesson1}.  Here we note that the procedure has been applied to
cosmologies of the Friedmann-Robertson-Walker (FRW) type \cite{coley}, 3D spherically symmetric
solutions \cite{acm2}, solutions with off-diagonal metrics \cite{leo2}, G$\ddot{\mbox{o}}$del-type
spacetimes \cite{crt}, and solutions containing a big bounce \cite{liuwes2,xlw,wlx,xuliu,likwes}.
General theorems have also been proven, having to do with the field equations \cite{leo3}, dynamics
\cite{bilsaj} and the algebraic classification of 5D solutions with their associated 4D
stress-energy tensors \cite{prt}.  However, there is the constraint that the $T_{\alpha\beta}$
given by algebra should correspond to the properties of matter indicated by observational cosmology.
For the early universe, this means that the equation of state for the matter should be close to that
of the ``classical vacuum''.  Here, the sum of the density $\rho$ and pressure $p$ is zero, as in
inflationary cosmology \cite{linde}.  We now proceed to this and other consequences of metric
(\ref{manifold}), to investigate its physical acceptability.

The line element (\ref{manifold}) can be written in the useful ``canonical'' form \cite{mlw} such that
\bea
dS^{2} = \frac{\ell^{2}}{L^{2}}\left[dt^{2}
         - \left[L\sinh\left(\frac{t}{L}\right)\right]^{2}d\sigma_{3}^{2}\right]
         - d\ell^{2}.
\label{canonical}
\eea
So with the 4D spacetime metric
\bea
g_{\alpha\beta} &=& \mbox{diag}\left[1,-\mathcal{F}_{k}(t,r),
-r^{2}\mathcal{F}_{k}(t,r),-r^{2}\sin^{2}\theta\mathcal{F}_{k}(t,r)\right]\nonumber\\
\mathcal{F}_{k}(t,r) &=& \left[L\sinh\left(\frac{t}{L}\right)\right]^{2}
                         \left(1+\frac{kr^{2}}{4}\right)^{-2},
\label{4dmetric}
\eea
we find the components of the stress-energy tensor $T_{\alpha\beta}=G_{\alpha\beta}$ to be
$T_{0}^{0}=3/L^{2}$, $T_{1}^{1}=T_{2}^{2}=T_{3}^{3}=3/L^{2}$.  In comoving coordinates this
defines an energy density $\rho=T_{0}^{0}=3/L^{2}$ and pressure
$p=-(T_{1}^{1}+T_{2}^{2}+T_{3}^{3})/3=-3/L^{2}$ for a vacuum with cosmological constant
$\Lambda=3/L^{2}$ and equation of state $\rho+p=0$.  The 4D Ricci scalar is
$R=R_{\alpha\beta}g^{\alpha\beta}=12/L^{2}$, and the 4D curvature scalar is
$K=R_{\alpha\beta\gamma\delta}R^{\alpha\beta\gamma\delta}=24/L^{4}$.  This latter scalar
implies that there are no singularities in the manifold because the constant $L\neq0$.

Let us now look at (\ref{manifold}), viewing its 4D part as describing an FRW model.  The
4D hypersurfaces $\ell=\mbox{constant}$ therefore describe cosmologies with scale factor
given by $\mathcal{S}=\mathcal{S}(t)=L\sinh(t/L)$.  Here the Hubble parameter
$H\equiv\dot{\mathcal{S}}/\mathcal{S}$ and deceleration parameter
$q\equiv-\mathcal{S}\ddot{\mathcal{S}}/\dot{\mathcal{S}}^{2}$ (with
$\dot{\mathcal{S}}=\tsent{\mathcal{S}}{t}$) are found to be
\bea
H = \frac{1}{L\tanh\left(\frac{t}{L}\right)}
\quad
\mbox{and}
\quad
q = -\tanh^{2}\left(\frac{t}{L}\right).
\eea
We note that $H$ is infinite at $t=0$ and goes to $1/L=\sqrt{\Lambda/3}$ as
$t\rightarrow\infty$, which is the Hubble parameter for de Sitter spacetime.  Also, $q$
starts at zero when $t=0$ and goes to $-1$ for $t\rightarrow\infty$.  This is in line with
astrophysical data which currently constrain the deceleration parameter to $-1\leq q\leq1$.
Thus we conclude that our solution (\ref{manifold}) describes an inflationary spacetime on
$\ell=\mbox{constant}$ hypersurfaces, where the vacuum has repulsive properties.

The preceeding paragraphs show that (\ref{manifold}) has physical properties consistent with
those of inflationary cosmology.  However, the motivating factor for the latter approach
to cosmology is that the (4D) horizon should grow fast enough to resolve certain problems
of astrophysical nature, primarily to do with the $3$ Kelvin microwave background
\cite{bps,wesson2}.  First, we recall that the horizon distance at time $t$ for any FRW model
can be defined \cite{peacock} such that
\bea
\int_{r}^{0}dr\left(1+\frac{kr^{2}}{4}\right)^{-1}
= \int_{0}^{t}\frac{dt^{\prime}}{\mathcal{S}(t^{\prime})}.
\eea
Multiplying both sides by the scale factor $\mathcal{S}(t)$ then gives
\bea
d_{PH} = \mathcal{S}(t)\int_{r}^{0}dr\left(1+\frac{kr^{2}}{4}\right)^{-1}
       = \mathcal{S}(t)\int_{0}^{t}\frac{dt^{\prime}}{\mathcal{S}(t^{\prime})},
\eea
which defines the proper distance to the particle horizon at time $t$.  For the spacetime
(\ref{4dmetric}) this is
\bea
d_{PH} = 2L\sinh\left(\frac{t}{L}\right)
         \left[\mbox{arctanh}(1) - \mbox{arctanh}\left(\mbox{exp}\left(\frac{t}{L}\right)\right)\right].
\eea
Here we see that $d_{PH}$ is infinite because $\mbox{arctanh}(1)$ is infinite.  This means that during
the inflationary period that the solution (\ref{manifold}) describes on $\ell=\mbox{constant}$
hypersurfaces, the entire universe is in causal contact.  This is in line with the apparent isotropy
of the microwave background.

Finally, we would like to point out an interesting coordinate transformation
of the solution (\ref{manifold}).  Recall that $\sinh(t)=(\mbox{exp}(t)-\mbox{exp}(-t))/2$,
which with the coordinate change $t\rightarrow tL$ in (\ref{manifold}) gives
\bea
dS^{2} = \ell^{2}dt^{2} - \frac{1}{4}\ell^{2}\left(\mbox{e}^{t} + k\mbox{e}^{-t}\right)^{2}d\sigma_{3}^{2}
         - d\ell^{2}
\quad
\mbox{with}
\quad
k = -1.
\label{manifold2}
\eea
This form of the metric resembles a solution noted by McManus \cite{mcmanus}.  For the solution
(\ref{manifold2}), all 4D physical quantities are the same as those calculated for the solution
(\ref{manifold}), but with the replacement $L\rightarrow\ell$.  The 4D spacetime contained in
(\ref{manifold2}) therefore still describes an inflationary vacuum with equation of state
$\rho+p=0$.  An important difference is that the 4D curvature scalar for (\ref{manifold2}) is
$K=24/\ell^{4}$, which implies that the spacetime in (\ref{manifold2}) has a singularity at the
point where $\ell=0$.  This is in contrast to the solution (\ref{manifold}), for which all
physical quantities of spacetime were calculated to be (finite) constants.  Evidently the
simple coordinate transformation $t\rightarrow tL$ casts (\ref{manifold}) into a form where the
vacuum evolves in accordance with how $x^{4}=\ell$ is determined by the extra component of the
geodesic equation.  This issue from the mathematical side has to do with whether we take the
whole of the 4D part of the 5D manifold as defining the geometry of spacetime, or whether we
take the 4D part of the 5D manifold without its prefactor.  This 5D issue resembles the 4D one
in scalar-tensor theory, where it manifests itself as a choice between what are commonly called
the Jordan and Einstein frames.  From the physical side, the choice has to do with how we
define spacetime as a 4D slice of a 5D manifold; and we suggest that since the two choices only
become differentiated over cosmological timescales, that it is essentially one of observation
to decide.

\section*{Acknowledgment}

This work was supported in part by the Natural Sciences and Engineering Research Council
of Canada (NSERC).



\end{document}